\documentclass[12pt]{iopart}
\pdfoutput=0
\usepackage{graphicx}
\usepackage{iopams}

\begin{document}

\title[Border in complex networks, B. A. N. Traven\c{c}olo et. al]{Border Detection in Complex Networks}

  \author{Bruno A. N. Traven\c{c}olo$^1$, Matheus Palhares Viana$^1$ and
  Luciano da Fontoura Costa$^{1,2}$}

  \address{$^1$ Institute of Physics of S\~{a}o Carlos - University of
  S\~{a}o Paulo\\ Av. Trabalhador S\~{a}o Carlense 400, Caixa Postal
  369, CEP 13560-970 \\ S\~{a}o Carlos, S\~ao Paulo, Brazil.}
  \address{$^2$  National Institute of Science and Technology for Complex
  Systems.}
  \ead{\mailto{luciano@if.sc.usp.br}}

\begin{abstract}
One important issue implied by the finite nature of real-world
networks regards the identification of their more external (border)
and internal nodes.  The present work proposes a formal and objective
definition of these properties, founded on the recently introduced
concept of node diversity.  It is shown that this feature does not
exhibit any relevant correlation with several well-established complex
networks measurements.  A methodology for the identification of the
borders of complex networks is described and illustrated with respect
to theoretical (geographical and knitted networks) as well as
real-world networks (urban and word association networks), yielding
interesting results and insights in both cases.
\end{abstract}

\maketitle

\section{Introduction}

Complex networks have progressed all the way from the initial
topological characterization of the Internet and WWW scale free
properties (e.g.~\cite{Barabasi1999,Faloutsos1999}) to becoming a
well-established and formalized research area
(e.g.~\cite{Barabasi2002Survey,Bocaletti2006Survey,Costa2007Survey})
with myriad of applications (e.g.~\cite{Costa2007AppSurvey}).  Yet,
given the relatively recent history of this field, there are still
several fundamental aspects which deserve further attention from the
complex networks community.  The current work addresses one of such
fundamental and largely overlooked aspects, namely the problem of
defining and identifying the \emph{borders} of several types of
networks.  We should make clear at the outset that there is no formal
definition of the borders of networks, so that their identification is
intrinsically related to the own definition of that concept.

\begin{figure}[ht]
\begin{center}
\includegraphics[width=0.8\linewidth]{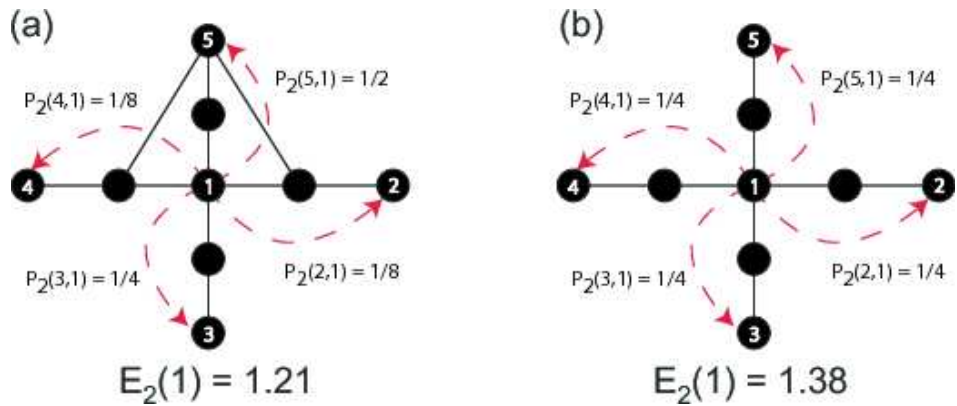}
\caption{Illustration of the concept of diversity entropy: (a) node
$1$ has low diversity entropy ($E_2(1)=1.21$) as a consequence of
its unbalanced transition probabilities to the nodes reachable after
two steps of a self-avoiding random walk; and (b) node $1$ has the
highest diversity ($E_2(1)=1.38$) considering four accessible
nodes.} \label{fig:dv_example}
\end{center}
\end{figure}

Recent works~\cite{Costa2008InOutAcc,Travencolo2008} have proposed
the use of the entropy of the transition probabilities between
nodes, with respect to a dynamics such as random walks and
self-avoiding random walks, in order to quantify the diversity of
access of individual nodes. The idea underlying such approach, as
several other entropy-based methodologies, is conceptually simple
and powerful. It is illustrated in figure~\ref{fig:dv_example} with
respect to a simple graph. Though in (a) node $1$ can access four
nodes after two steps along a self-avoiding random walk, the
transition probabilities to each of these nodes is markedly
different.  In the situation illustrated in
figure~\ref{fig:dv_example}(b), node $1$ can also access four nodes,
but with equal transition probabilities. The effectiveness of access
from node $1$ to the other nodes can be nice and effectively
quantified in terms of the entropy of the transition probabilities
to the accessible nodes~\cite{Travencolo2008}, implying the
situation in (b) to be much more balanced and effective than the
situation in (a). Actually, it can be showed that the minimal time
for accessing all the nodes reachable from a given reference node
after $h$ steps is minimal when the entropy is maximum.  Such a
basic principle of the diversity entropy concept is adopted in the
present work in order to define the borders of networks, in the
sense that the diversity would be directly related to the
internality of the nodes.

Though the relationship between diversity entropy and the
internality of the nodes had been hinted
previously~\cite{Costa2008InOutAcc,Travencolo2008}, the present work
reports a systematic investigation of such a definition of the
borders of networks from several perspectives while taking into
account model (theoretical) and real-world networks. After formally
defining the concept of internality of each individual node in any
network, we provide some analytical motivation with respect to
diffusion (i.e. random walks) in regular lattices.  Then we focus on
geographical networks, whose borders can be intuitively related to
their geographical structure (everybody has a conceptual idea of the
borders of a city, for example).  It is shown that the geographical
borders tend to be in close agreement with the topological borders
identified by the diversity entropy for both 2D and 3D geographical
structures. Then, we investigate how the concept of internality
extends to non-geographical networks with respect to knitted
networks~\cite{Costa2007Knitted}, which are regular but
non-geographical stochastic structures.  We perform this study by
making `holes' in the original networks, so that the borders of the
holes can be well-defined topologically as corresponding to the
limits of the holes.  We show that the diversity entropy approach
can precisely identify the original borders created by the holes.
Subsequently, we take into account rewired (by using the procedure
described by Maslov and Sneppen in~\cite{Maslov2002}) versions of
geographical networks, which allows us a series of insights on how
the borders of initially geographically-constrained networks change
with perturbations.  The important issue of quantifying the degree
of possible correlations between the diversity entropy and
well-established measurements such as the degree and betweenness
centrality are also investigated. The results corroborate little
interrelationship between diversity entropy and such measurements,
implying that the former feature is indeed providing additional
information about the structure of the network. We conclude the
current work by analyzing real-world networks, namely the plant of
the town of S\~ao Carlos (SP, Brazil) as well as the network of word
associations in Lewis Carroll's Alice's Adventures in Wonderland.
Remarkable results are obtained regarding both cases, including the
identification of modules of internality in the town and centrality
of words in Alice's Adventures in Wonderland.

\section{Materials and Methods}

\subsection{Basic Concepts}

{\bf Complex Networks} An unweighted and undirected complex network
is a set of $N$ nodes linked by $E$ edges
\cite{Barabasi2002Survey,Costa2007Survey} that can be described by
the adjacency matrix $A_{i,j}$. This matrix has binary elements
$a_{i,j}$ that represent the presence or absence of connection
between each pair of nodes. When $a_{i,j}=1$ there is an edge
between the node $i$ and the node $j$. Otherwise, $a_{i,j}=0$
indicates that there is no connection between them. The degree of
the node $i$, $k_i$, is the number of nodes directly connected to
$i$. This quantity is related with the adjacency matrix by the
relation $k_i=\sum_{j=1}^Na_{i,j}$.

A \emph{walk} of length $h$ over the network is defined by a subset
of $h+1$ adjacent nodes (i.e., nodes that share at least one edge).
Considering a discrete time Markovian process over the network, the
\emph{transition probability} $P_h(j,i)$ is the probability that an
agent departing from node $i$ reaches the node $j$ after $h$ steps.
This probability can vary accordingly to the type of walk that is
adopted. In the case of \emph{self-avoiding random walks}, the
moving agent cannot repeat nodes or edges during the
walk~\cite{Costa2008InOutAcc,Travencolo2008}, while in traditional
\emph{random walks} the agent has no restriction to perform the walk
\cite{Noh2004RandonWalk}. We used self-avoiding random walks for all
the examples presented in this paper, as with this dynamics it is
possible to reach more nodes in fewer steps.

{\bf Diversity Entropy} The diversity entropy of a particular node
$i$ measures how diverse is the access from this node to the other
nodes of the network through a walk of length $h$. Let $\Omega$ be
the set of all nodes of the network, except $i$, and $N$ be the
number of nodes of the network. The normalized diversity entropy
$E_h(\Omega,i)$ can be expressed as~\cite{Costa2008InOutAcc,
Travencolo2008}

\begin{equation}
    E_h(\Omega,i) = -\frac{1}{log(N-1)} \sum_{j=1}^{N}
    \left\{
        \begin{array}{rll}
            P_h(j,i)log(P_h(j,i))  & \textrm{if $P_h(j,i) \neq 0$} \\
            0 & \textrm{if $P_h(j,i) = 0$} \\
        \end{array}
    \right.
    \label{eq:diversity}
\end{equation}

\subsection{Border Detection}

The concept of diversity is intuitively related to the definition of
the border of a network. The main idea behind this property is that
the peripheral nodes have not many options for proceeding a random
walk other than accessing the internal nodes of the network,
resulting in low diversity values~\cite{Travencolo2008}. On the
other hand, non-border nodes tend to have more effective and
balanced access to the most part of the internal and peripheral
nodes of the network, resulting in high diversity entropy values. In
order to demonstrate this property analytically, we resourced to a
bidimensional grid, as explained below.

\subsubsection{A theoretical approach}

\begin{figure}[ht]
    \begin{center}
        \includegraphics[scale = 0.5]{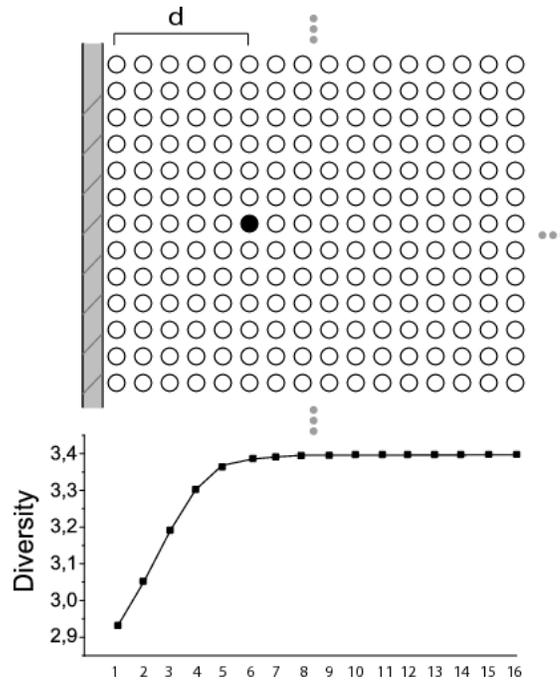}
        \caption{Illustration of the border effects on diversity entropy.
        (a) A semi-infinite grid network with border. The central node
        is painted in black and $d$ is the distance between the
        central node and the border. (b) Diversity entropy as function
        of the distance $d$ from the border.}
    \label{fig:result_theory}
    \end{center}
\end{figure}

Let us consider a bidimensional semi-infinite grid illustrated in
figure~\ref{fig:result_theory}. We want to investigate the diversity
entropy of the central node (shown in black in
figure~\ref{fig:result_theory}) located at distance $d$ from the
border of the grid. We start by evaluating the number of random
walks of length $h$ between the central node and the node at the
coordinate $(x,y)$, which is given by:

\begin{equation}
    W_h(x,y) = {\cal C}(h,x){\cal C}(h,y)-\beta_h(x,y)
\end{equation}

where ${\cal C}$ denotes the binomial symbol and

\[\beta_h(x,y) =
\left\{
    \begin{array}{l}
            {\cal C}(h,d+x+1){\cal C}(h,d+y+1)\\
            \;\;\;\;\;\quad \mbox{if $x,y\leq h-d-1$ and $x+y\geq h-d-1$}\\
            0\\
            \;\;\;\;\;\quad \mbox{if $x,y\leq h-d-1$ and $x+y < h-d-1$}
  \end{array}
\right.
\]

The probability that a walker reaches the node at coordinate $(x,y)$ after $h$ steps
starting at the central node is

\begin{equation}
    P_h(x,y) = \frac{W_h(x,y)}{\sum_m\sum_nW_h(m,n)}
\end{equation}

Then, the diversity entropy of the central node is

\begin{equation}
    E_h =
    -\frac{1}{\log(N-1)}\sum_x\sum_yP_h(x,y)\log\left(P_h(x,y)\right)
\end{equation}

Observe that the sums run over all coordinates for which the value of
$P_h(x,y)$ is not null. Figure~\ref{fig:result_theory}(b) shows the
diversity entropy as function of the distance $d$ of the central node
from the border, considered a fixed walk length step of $h=7$. Note
that in this figure the value of diversity entropy is lower near the
border and constant when $d>h$ (i.e. the central node does not suffer
border effects).

\subsection{Border Detection Using Other Measurements}

In addition to the diversity entropy, we also evaluate the use of
other measurements in order to define de border of networks. The
considered features include: the node degree, the average shortest
path length (\textit{ASPL}), and betweenness centrality ($B$). The
ASPL for the node $i$ is the average length of all shortest path
between $i$ and all other nodes of the network
\cite{Costa2007Survey}. The betweenness centrality of the node $i$ is
defined as $B_i =
\sum_{i\neq p\neq q}\sigma_{p,q}(i)/\sigma_{p,q}$, where
$\sigma_{p,q}$ is the number of shortest paths from $p$ to $q$ and
$\sigma_{p,q}(i)$ is the number of shortest path from $p$ to $q$ that
go through $i$~\cite{Costa2007Survey}.

\subsection{Datasets}

In order to evaluate the efficiency of the proposed methodology, we
used three hypothetical geographical networks with different shapes
and physical constraints, such as bottlenecks and internal holes.
Geographical networks were used as they are the only type of network
where the border nodes can be defined visually, allowing immediate
evaluation of the accuracy of the border-detection methodology.

Figure~\ref{fig:result_SelfAvoiding} shows the geographical networks
adopted in our analysis. The first network, depicted in (a),
consists of a single square-shaped network. The case (b) regards a
network that includes bottlenecks (narrow connections) as well as
internal holes. The third case, shown in (c), shows a sphere-shaped
three-dimensional geographical network. In all cases, the nodes were
generated by sampling spatial points in a space constrained by the
bounding desired physical shape (e.g. points chosen within a circle
of a given radius). The edges of the network were subsequently
defined by Delaunay triangulation of the sampled points.

In order to validate the methodology for non-geographical networks, we
performed the border detection over the Knitted model
\cite{Costa2007Knitted}. This network is built from a set of non
connected nodes labeled from 1 to $N$. The basic step while growing
this type of network involves shuffling the labels in an arbitrary
sequence and defining the edges by connecting consecutive labels in
that sequence. Note that the last label in the sequence is connected
to the first one. This step is repeated $n$ times. As result of this
process, we obtain a non-geographical network where all nodes have
degree $k=2n$.

In addition to the hypothetical networks described above, we also
considered two real-world examples of networks: the first refers to
the network of urban streets of the Brazilian town of S\~ao Carlos.
In this network, each node represents a street crossing or beginning
of routes, while the edges represent the streets
(see~\cite{Travencolo2008} for additional information); the second
is a word association network, built considering the relationship
between words in the Lewis Carroll book Alice's Adventures in
Wonderland. The nodes of this network represent the words, while the
edges link the words that are adjacent in the
text~\cite{Antiqueira2007}.

\section{Results}

This section presents six different results of the border detection
methodology considering different measurements and data. First, the
border of the geographical networks presented in
figure~\ref{fig:result_SelfAvoiding} was detected in terms of the
diversity entropy and self-avoiding random walks dynamics. Next, the
results of the proposed methodology applied to non-geographical
network are presented. In another analysis, we considered the border
detection of a network four times denser than one of the networks of
the first analysis. The next case shows an example of a geographical
network that was progressively rewired and, again, the diversity
entropy was used for border detection. Subsequently, the results of
border detection using other measurements (node degree, betweenness
centrality, and average shortest path length) are shown and
discussed. Finally, the results of border detection with respect to
the two real-world networks are presented and interpreted.

\subsection{Detected borders using diversity entropy}

\begin{figure}[ht]
\begin{center}
\includegraphics[width=\linewidth]{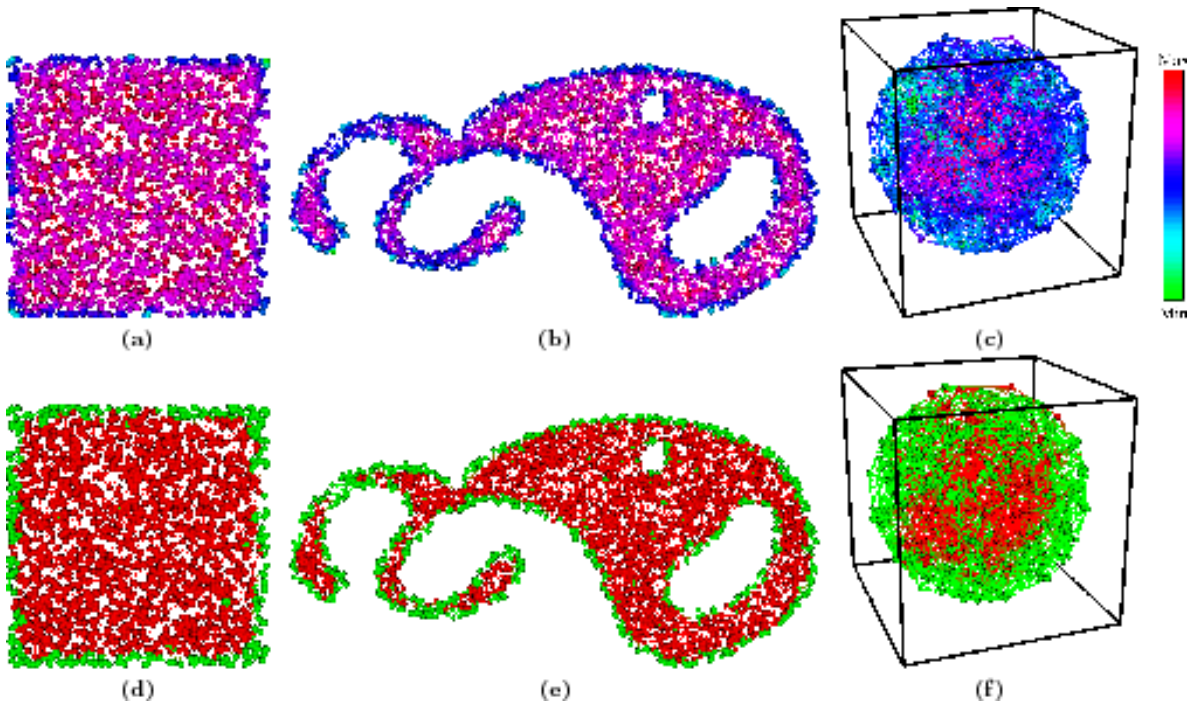}
\caption{Border detection using diversity entropy with self-avoiding
walks. (a-c) Three geographical networks used to evaluate the
methodology. Each node is colored with respect to their diversity
entropy value considering self-avoiding random walks and $h=3$. Note
that the network in (c) is three-dimensional.(d-f) Results of the
detection of the border (green nodes) by thresholding the diversity
entropy.} \label{fig:result_SelfAvoiding}
\end{center}
\end{figure}

The diversity entropy for the networks previously introduced were
calculated considering five steps ($h=1..5$). The nodes of the
networks shown in figure~\ref{fig:result_SelfAvoiding}(a-c) were
colored accordingly to their respective diversity entropy value for
$h=3$. The classification of a node as border or non-border can be
performed by choosing a threshold of the diversity at specific step.
The nodes whose diversity entropy is below the threshold value are
classified as border nodes.
Figure~\ref{fig:result_SelfAvoiding}(d-f) shows the detected border
nodes (green nodes). Notably, in all analyzed networks the border
nodes were successfully identified.

In addition, it is important to note that the thresholds used to
obtain these results were automatically chosen. The following
procedure was adopted: first, the minimum distance from each node to
the contour of the shape (the physical border) was determined. The
nodes nearest the border (i.e., have low distance values) were
classified as border, while the others were understood as non-border
nodes. Next, this classification was compared with the
classification obtained by performing many consecutive thresholds of
the diversity entropy value. The threshold value which result in the
best relation between correctly classified border/non-border nodes
was then used to define the border.

\subsection{Extension for non-geographical networks}

In this section we give some insights about the extensibility of the
diversity to detect borders in non-geographical networks. Unlike the
case for geographical networks, the definition of border of
non-geographical networks is not intuitive nor can be visualized. In
order to address such networks, we development a methodology to
assess the accuracy of the border detection methodology which
basically consists in creating holes in a network. The main idea is
to select a few nodes of the network and remove these nodes and
their neighbors up to some pre-defined distance. By doing so, some
holes in the network are created, and, as consequence, the neighbors
of the removed nodes can be considered as border nodes. Then, after
the identification of these border nodes, it is possible to evaluate
the proposed border detection methodology by thresholding the
diversity entropy and comparing the detected border with the actual
borders defined by the holes.

\begin{figure}[ht]
\begin{center}
\includegraphics[width=\linewidth]{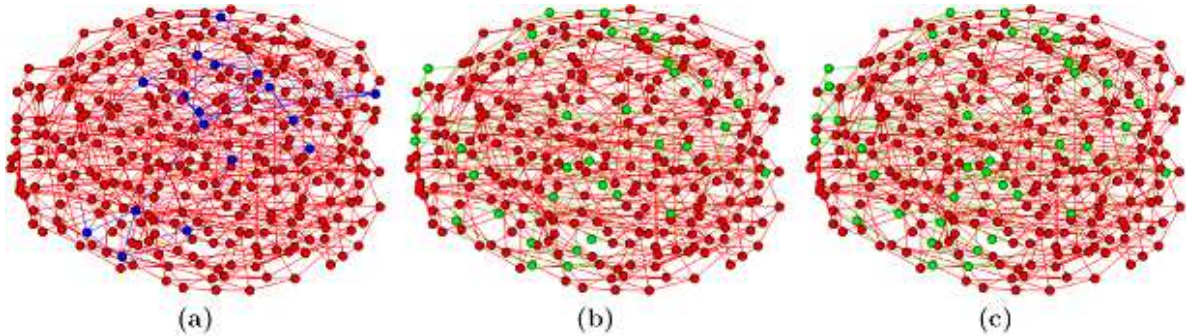}
\caption{Border detection on a knitted network. (a) Original network
where the blue nodes are removed in order to insert a hole on the
network. (b) Network with a hole. The green nodes represent the
border of this network. (c) Border detection using diversity entropy
(green nodes represent the border nodes). Note that the obtained
border is quite similar to the actual borders shown in (b).}
\label{fig:result_knitted}
\end{center}
\end{figure}

This process is illustrated in figure~\ref{fig:result_knitted}. A
knitted network with $N=300$ and $E=600$ is shown in
figure~\ref{fig:result_knitted}(a). The hole in this network was
obtained by removing the node with the highest diversity entropy
value (note that other criteria could also be used) and its
neighbors up to distance two. The border defined by this process is
shown in figure~\ref{fig:result_knitted}(b) (green nodes), while the
borders defined by thresholding the diversity entropy is shown in
figure~\ref{fig:result_knitted}(c). Note that the results are quite
good, with only a few misdetected nodes (4 nodes). For this
analysis, the threshold was automatically chosen in a procedure very
similar to that used in the geographical networks, with the
difference that instead of using the distance from the physical
border to define the actual border nodes, the nodes next to the hole
were defined as the actual border.

In addition to the analysis of the border detection on a knitted
network, we also tried the same procedure on a Barab\'asi-Albert
network (BA)~\cite{Barabasi2002Survey}. However, the obtained
results for a series of experiments showed that the ratio between
correctly detected border/non-border were not satisfactory. The main
reason for this result is that the average shortest path length in
these networks is usually very low (e.g., this value is only 7 in a
network with $N=1,000,000$ and $<k>=4$). As a consequence, the vast
majority of the nodes of the network will be very close to the
border defined by the holes, and their diversity entropy will be
reduced (due to the proximity to the border). This reduction
increases substantially the chances of misclassification. It is
important to stress that the hole approach being unsuitable to
evaluate the border detection methodology on BA networks does not
invalidate our methodology. Given the other results reported in this
work, as well as the own motivation for defining the borders as
nodes of low diversity, it is reasonable to expect that the method
will also correctly identify the borders in small-world networks
such as Barab\'asi-Albert.

\subsection{Border detection for a denser network}

\begin{figure}[ht]
\begin{center}
\includegraphics[width=\linewidth]{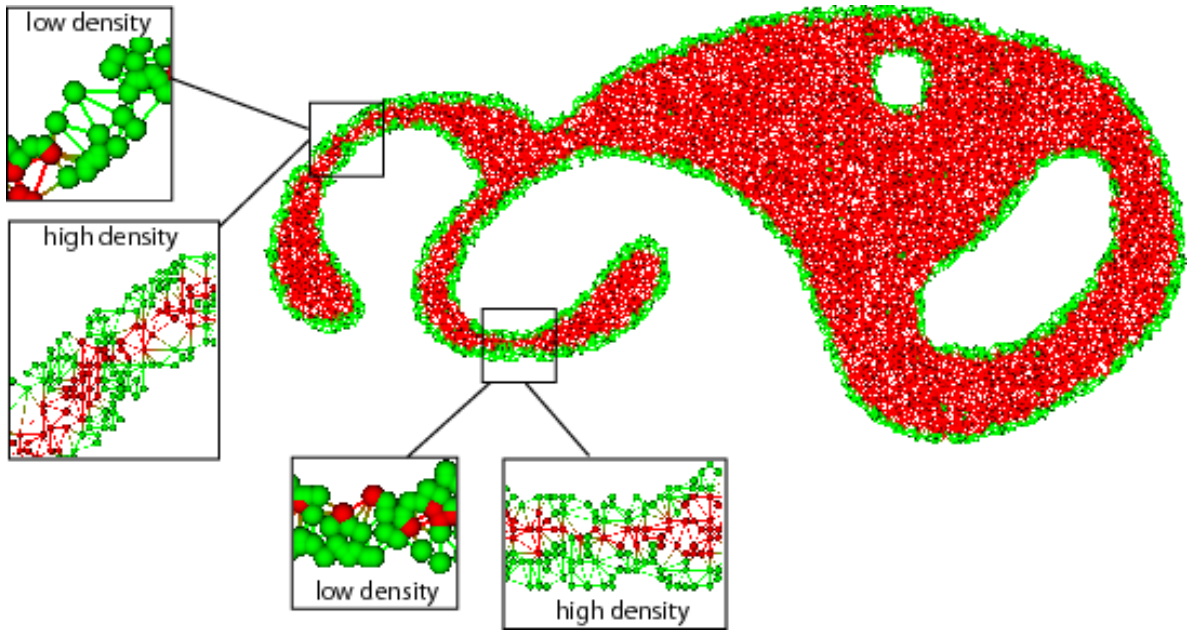}
\caption{Border detection for a denser network. This network has the
same shape of the network shown in
figure~\ref{fig:result_SelfAvoiding}b, but it is four times denser.
The border nodes were detected (green nodes) and, when compared with
the results obtained for the less dense network
(figure~\ref{fig:result_SelfAvoiding}f), it can be seen that the
obtained  results are quite similar. Note that, as shown by the
zoomed regions, the main differences between these two networks can
be found in the narrow regions, associated to smaller spatial and
topological scales.} \label{fig:result_Density}
\end{center}
\end{figure}

This section reports a comparison of the results of the proposed
border detection methodology considering networks with the same
shape but with different density of nodes. For this purpose, we
resourced to networks that contain narrow regions and holes
(figure~\ref{fig:result_SelfAvoiding}b), as well as a version of the
same networks four times denser (with reference to nodes per square
unit), shown in figure~\ref{fig:result_Density}a. For the latter
case, the border detection methodology was applied, i.e., the
diversity entropy considering self-avoiding walks was calculated and
a threshold on the diversity value was used to define the border
(green nodes in figure~\ref{fig:result_Density}).

The main difference between the results obtained for the two
considered networks refer to the narrow regions of the shapes. The
zoomed squares in figure~\ref{fig:result_Density} show some of these
regions for both networks. While most of the nodes situated in
narrow regions in the less dense network are considered as border,
in the denser network only the most peripheral nodes of these
regions are classified as borders. This result shows that the method
is sensible to the spatial resolution of the nodes of the
geographical network.

\subsection{Border detection in a rewired network}

\begin{figure}[ht]
\begin{center}
\includegraphics[width=\linewidth]{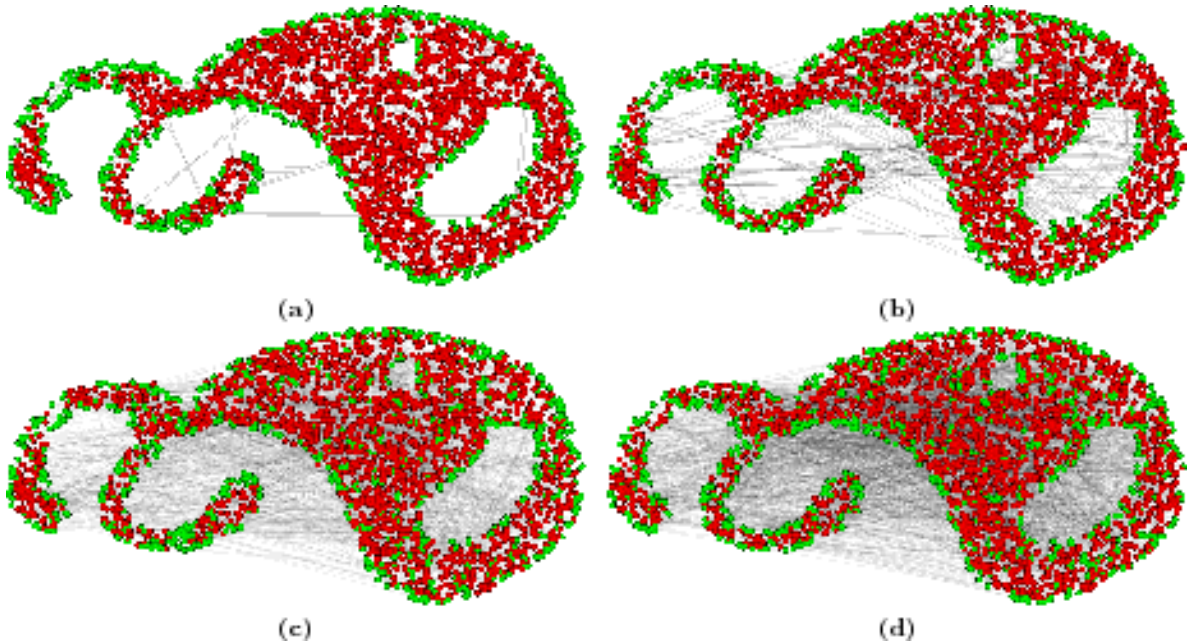}
\caption{Border detection in a rewired network. The network shown in
figure~\ref{fig:result_SelfAvoiding}b had 0.1\%~(a), 1\%~(b),
5\%~(c) and 10\%~(d) of its edges rewired and the border was
detected (green nodes). (a) Due to the low percentage of rewiring,
no significant changes were detected during border detection. (b-d)
When considering higher percentage of rewiring, a dislocation of the
border to the internal regions of the networks can be clearly
identified. } \label{fig:result_rw}
\end{center}
\end{figure}

In this section we address the results of the border detection
methodology when applied to rewired versions of geographical
networks. Note that, when using rewired versions of a network it is
reasonable to consider that part of the properties of the original
(non-rewired) network are maintained, i.e., some regions of the
rewired networks will still have the properties implied by the
physical constraints while others will acquire non-geographical
behavior (e.g. emergence of small world effect). Therefore, it is
expected that the border detection methodology applied to these
rewired network will not consider as border some of the former border
nodes and additionally it will classify as border some of the former
internal (non-border) nodes.

In order to evaluate this process, four different rewired versions
of the network presented in figure~\ref{fig:result_SelfAvoiding}b
were analyzed. Figure~\ref{fig:result_rw} shows the obtained
networks with 0.1\%~(a), 1\%~(b), 5\%~(c) and 10\%~(d) of rewired
edges. Note that, in this figure, the nodes are already colored
accordingly to the border/non-border classification. The threshold
criteria used to define the border in all cases was the same used
previously, i.e., the configuration which best approximates the
physical border of the original network. That is the reason why many
of the detected border nodes still remain near of the physical
border.  Particularly, for the first case, (a), the very low
percentage of rewiring did not change the properties of the network
and, as a consequence, the obtained result is quite similar to the
original, non-rewired network. On the other hand, for the three
remainder cases, the classification of the nodes as border nodes
included some former internal nodes and excluded some former border
nodes. Note that this result can also be understood as an extension
of the proposed methodology to non-geographical networks.

\subsection{Detected borders using node degree, betweenness
centrality and average shortest path length}

\begin{figure}[ht]
\begin{center}
\includegraphics[width=\linewidth]{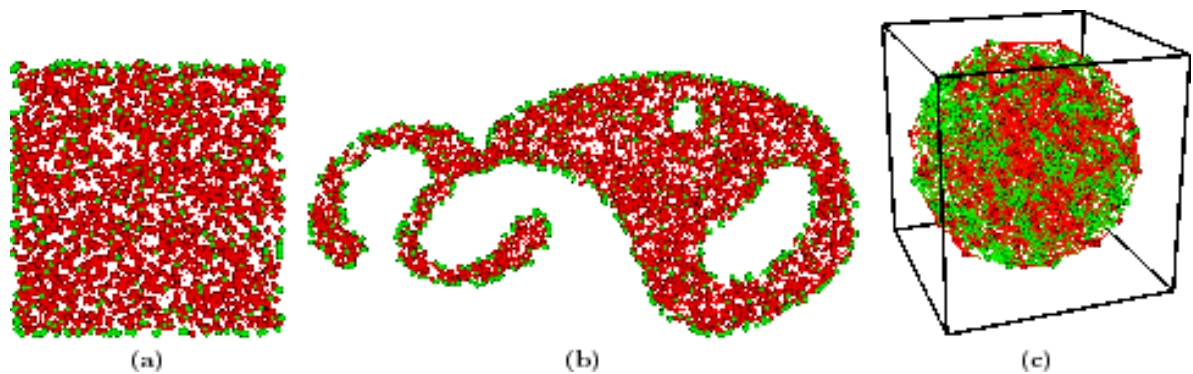}
\caption{Border detection using the node degree. Although the border
was properly detected, many internal nodes were also considered as
border.}\label{fig:result_nviz}
\end{center}
\end{figure}

\begin{figure}[ht]
\begin{center}
\includegraphics[width=\linewidth]{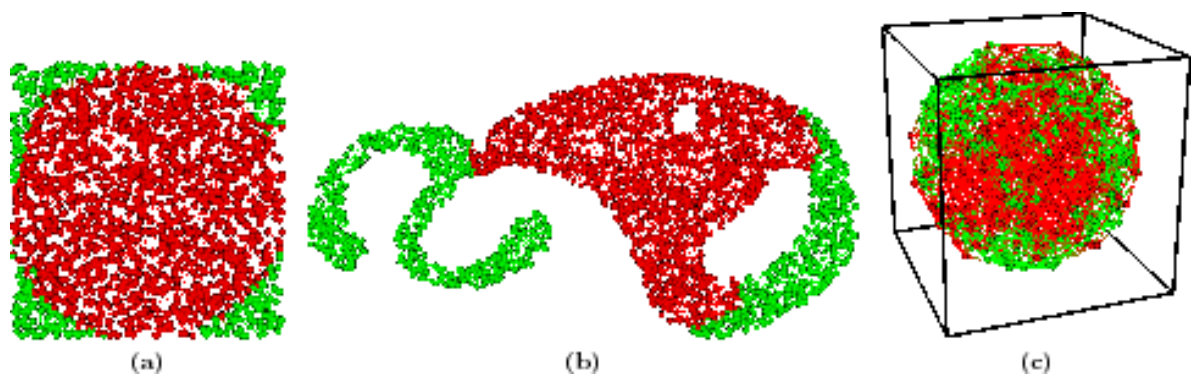}
\caption{Border detection using the average shortest path length. In
contrast with the diversity entropy, many nodes were
misclassified.}\label{fig:result_asp}
\end{center}
\end{figure}

\begin{figure}[ht]
\begin{center}
\includegraphics[width=\linewidth]{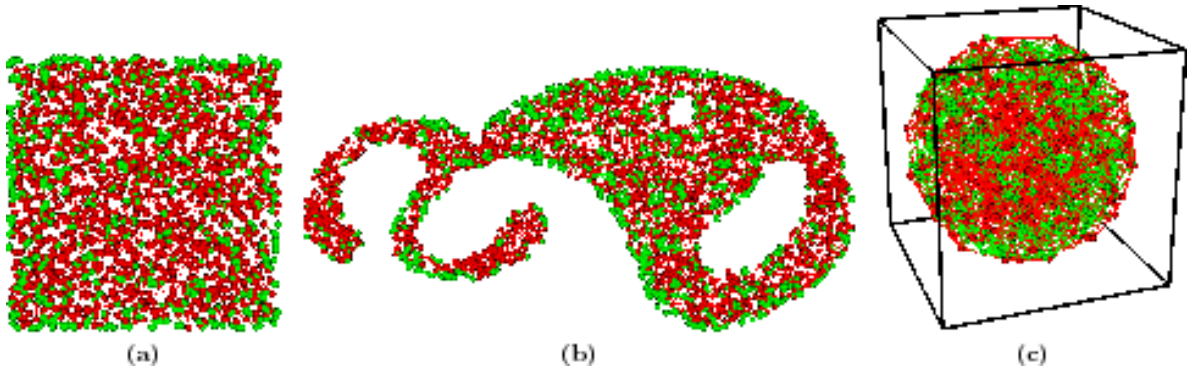}
\caption{Border detection using the betweenness centrality. In this
case, the noise level is high, i.e. the amount of internal nodes
classified as border is high. This is a consequence of the fact that
the betweenness centrality can assume low values even for the nodes
located in the internal region of the
network.}\label{fig:result_bcentrality}
\end{center}
\end{figure}

The results of the border detection obtained by using the node
degree, the betweenness centrality and the average shortest path
length are shown in the
figures~\ref{fig:result_nviz},~\ref{fig:result_asp},
and~\ref{fig:result_bcentrality}. For all cases, the best threshold
value was chosen in the same way described previously. Among these
three measurements, the best result was found for the node degree.
In this case, the border was properly detected at the expense that
many internal nodes, which are not border, were classified as
border. The same issue occurs, in a worse manner, when the
betweenness centrality is used to detect the borders. Finally, when
using the ASPL to detect the borders, it is clear from
figure~\ref{fig:result_bcentrality}(a-b) that only the nodes located
far from the geographical center of the network are classified as
border, resulting in a completely incorrect result. In
figure~\ref{fig:result_bcentrality}(c), the detection of the border
presented a similar result with respect to the result obtained by
using the diversity entropy.

In order to investigate the relationship between these three
measurements and the diversity entropy, Table~\ref{tab:correlations}
shows their Pearson correlation coefficients. It can be noted from
this table that no significant correlation was found with respect to
any of the three considered networks, except for the network from
figure\ref{fig:result_SelfAvoiding}(c) (3D sphere), whose diversity
entropy presented high correlation with the ASPL.

\begin{table}
\begin{tabular}{|l|r|r|r|}
  \hline
  Network  & DV x Degree & DV x BC & DV x ASPL
  \\ \hline
  Figure~\ref{fig:result_SelfAvoiding}(a) & 0.73 & 0.56 & -0.59  \\
  Figure~\ref{fig:result_SelfAvoiding}(b) & 0.76 & 0.18 & -0.42 \\
  Figure~\ref{fig:result_SelfAvoiding}(c) & 0.76 & 0.72 & -0.91 \\
  \hline
\end{tabular}
\caption{Pearson correlation coefficients. DV: Diversity entropy;
ASPL: Average shortest path length; BC: Betweenness
centrality.}\label{tab:correlations}
\end{table}

\subsection{Real-world examples}

In order to best illustrate the potential of the proposed methodology,
in this section we present the results considering two real-world
network examples: the geographical network of the urban streets of
S\~ao Carlos and the word association network derived from the book
Alice's Adventures in Wonderland.

\begin{figure}[ht]
\begin{center}
\includegraphics[scale=1]{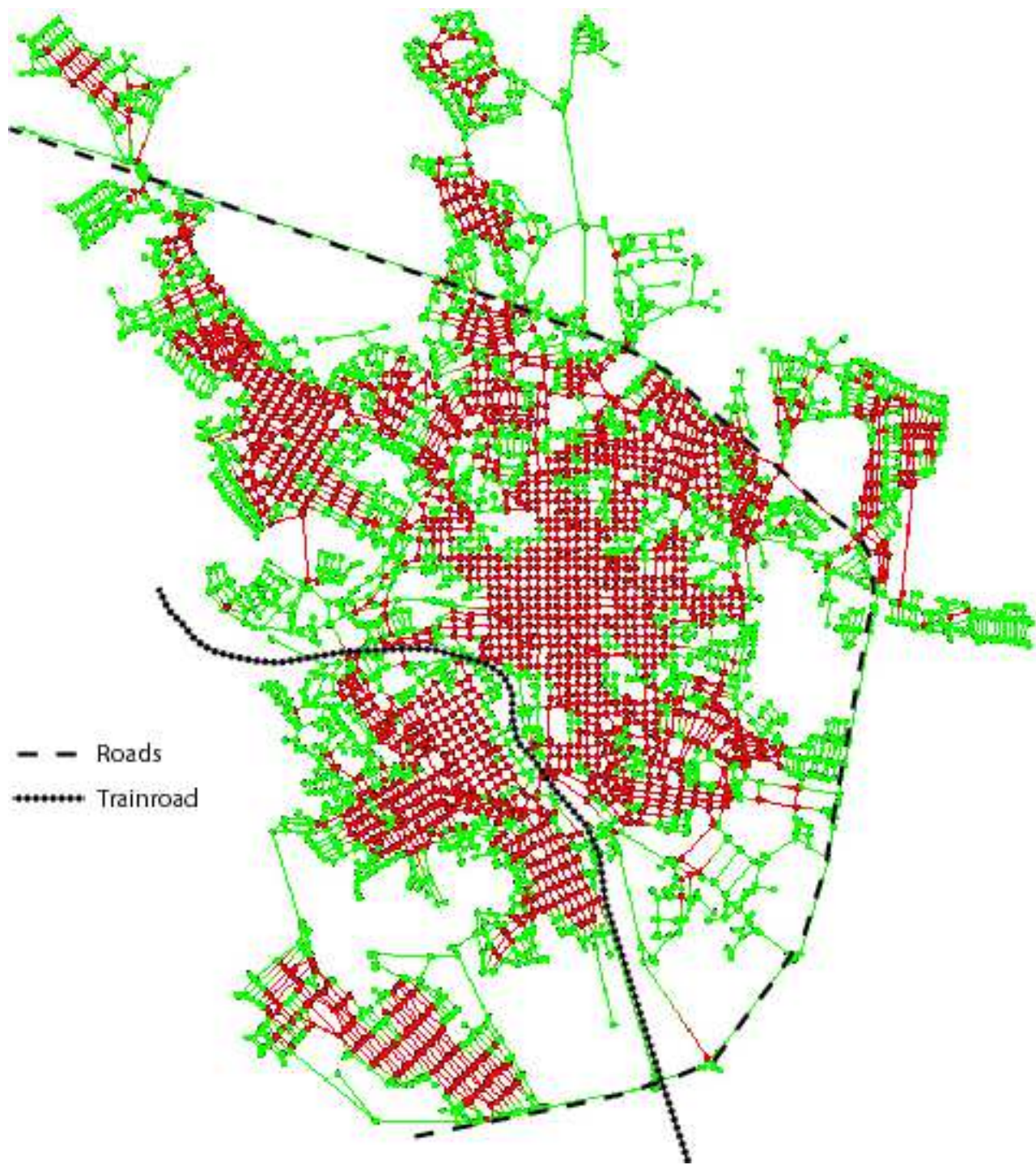}
\caption{Border detection in the network of urban streets of the
town of S\~ao Carlos - SP - Brazil. The border is indicated by the
green nodes. Note that the highway and railroad that cross this town
also defined specific borders which were properly identified by the
diversity methodology.}\label{fig:sanca}
\end{center}
\end{figure}

The network of the streets of S\~ao Carlos has $N=4537$ nodes and
$E=7527$ edges. The diversity entropy was estimated for $h=1..5$. In
order to determine the borders of this network, the threshold of the
diversity entropy was manually chosen. The obtained result is shown
in figure~\ref{fig:sanca}, where the green nodes represent the
border nodes. This figure also shows the railway and the highways
that cross the urban perimeter of the city. These structures impose
important physical constraints in the planning of the streets of the
city, slicing the urban area and giving rise to many border nodes
along these ways.  Observe that such border nodes, though not being
directly related to the periphery of the network, were all properly
detected by the proposed diversity methodology.

\begin{figure}[ht]
\begin{center}
\includegraphics[width=\linewidth]{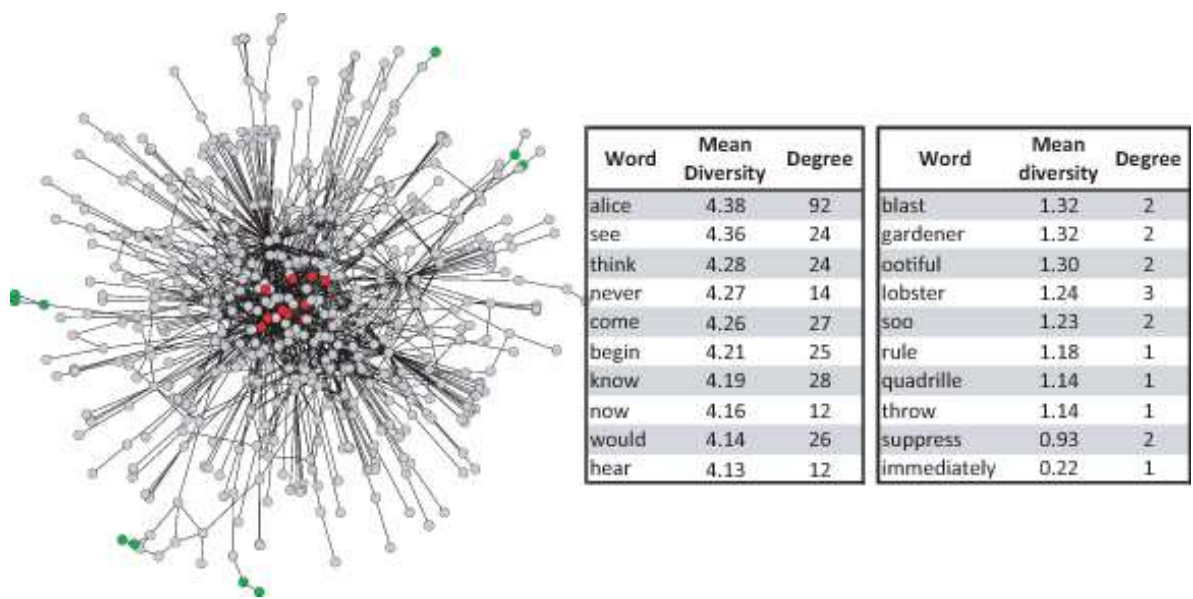}
\caption{Network derived from words relationship in the book Alice's
Adventures in Wonderland. The green and red nodes represent the
lowest and highest values of diversity entropy, respectively. The
words corresponding to these nodes, as well their diversity entropy
and degree, are shown in the side tables. Interesting, the node with
highest degree (130), which correspond to the verb \textit{say},
does not appear in the above list.}\label{fig:alice}
\end{center}
\end{figure}

The second real-world example analyzed, the word association network
from the book Alice's Adventures in Wonderland, has $N=1929$ nodes
and $E=9290$ edges. This network was built considering the words as
nodes and by linking adjacent words in order to define the edges. To
reduce the size of the network, the words were lemmatized (i.e.
reduced to their canonical form) and the stop-words (e.g. the verb
to be, prepositions, conjunctions, pronouns and articles) were not
considered. In addition, we excluded the nodes of the network which
had less than two connections and selected the biggest connected
component of the remainder network. The resulting network, which has
$N=505$ and $E=1102$, is shown in figure~\ref{fig:alice}. The
diversity entropy was computed for this network ($h=1..5$) and the
ten lowest and highest values, together with their corresponding
word, are shown in figure~\ref{fig:alice}. It is clear that the most
internal words (corresponding to the most internal nodes) seem to
correspond to the most important terms in the book, while the border
words tend to present a secondary nature regarding the main subjects
in this book. In this sense, it is possible that the diversity
quantifies in some way the centrality of words and concepts in
texts.

\begin{figure}[ht]
\begin{center}
\includegraphics[width=\linewidth]{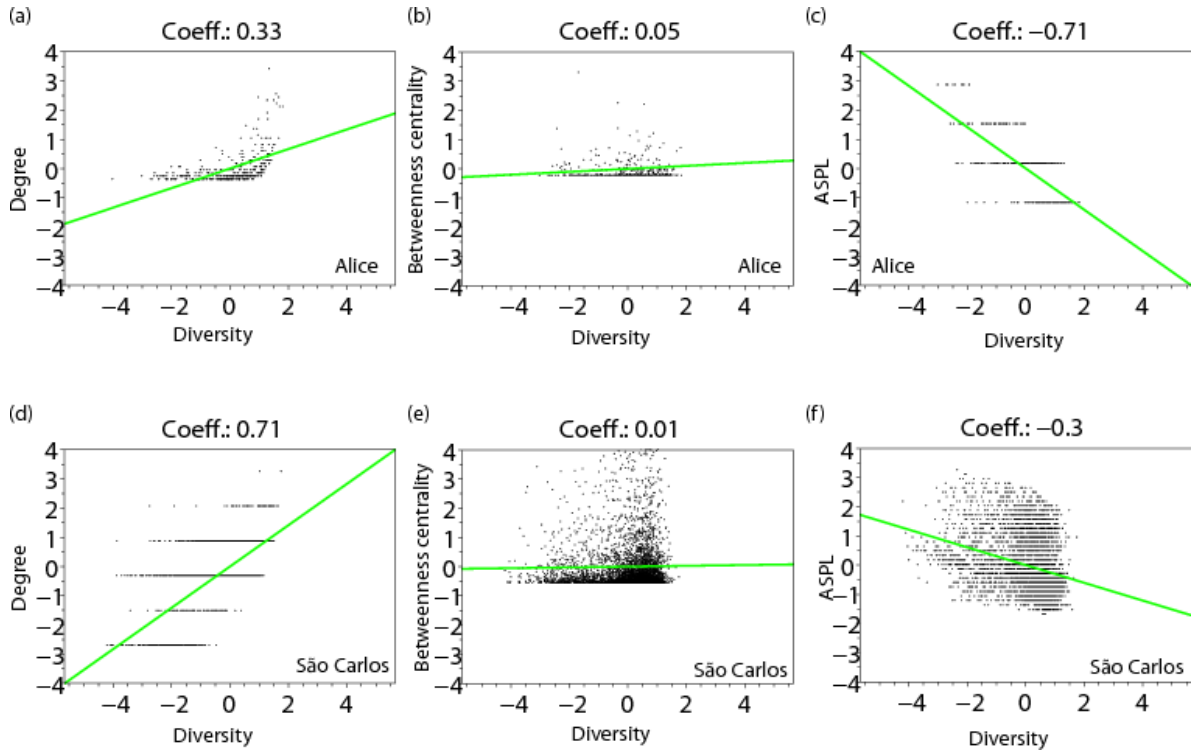}
\caption{Pearson correlation coefficients between the diversity
entropy and the number of neighbors (a,d), the between centrality
(b,e) and ASPL (c,f) considering the word association network (a-c)
and the streets network (d-f). Note that no considerable correlation
between these measurements was identified for these networks
(Coeff.: Pearson correlation coefficient).}\label{fig:correlations}
\end{center}
\end{figure}

Figure~\ref{fig:correlations} shows the Pearson correlation
coefficients obtained for the above two real networks considering
the diversity entropy and the other three previously introduced
measurements: degree, betweenness centrality, and ASPL. The obtained
results indicate that there is no strong correlation between these
measurements.

\section{Conclusions}

Despite all the current investigations in complex networks research,
some important related issues have received little attention.  Given
that real-world networks are necessarily finite, one important
aspect concerns the definition and identification of their borders,
a concept immediately related to internality/externality of nodes.
The current work has addressed these important problems.  More
specifically, we used the recently introduced concept of diversity
(e.g.~\cite{Costa2008InOutAcc,Travencolo2008}) in order to quantify
the potential of accessibility from and to each node while
considering a specific dynamics, in the present case self-avoiding
random walks. Nodes with more balanced transition probabilities to
other nodes, expressed in terms of entropy, are understood as being
more internal to the network, while the other nodes are associated
to the network borders.

The concept of borders is immediate and intuitive in geographical
networks, where the more internal nodes are usually found at the
geometrically more internal regions.  Therefore, we give special
attention to this type of networks in order to motivate and validate
our approach. Interestingly, the application of the diversity
methodology to geographical networks identified as borders not only
the nodes at the peripheral regions, but also those nodes which are
geographically more internal but are near geographical
discontinuities slicing the network.  The definition of borders in
non-geographical networks has been an important open question. We
showed that it is possible to extend the diversity approach to
define and identify borders in that type of networks.  We also
compared the diversity approach with methods founded on other
measurements such as degree, betweenness centrality and average
shortest path lengths.  None of these alternative approaches turned
out not to be able to properly detect the borders, even in the case
of geographical networks.  We also found no significant correlation
between the diversity and those measurements, which further
corroborates that that feature does provide additional information
about the topology of the networks. The potential of the methodology
proposed for the identification of the borders was illustrated with
respect to both theoretical (i.e.\ regular, geographical and knitted
networks) and real-world networks (i.e.\ the urban network of the
town of S\~ao Carlos and Carroll's Alice's Adventures in
Wonderland).  Several findings and insights were yielded by the
adopted approach, including the impressive performance for
identification of the borders in knitted networks, a
non-geographical structure.  In addition, it was showed that the
urban network presents borders not only at its periphery, but also
along the railway line and highway that happen to cross that town.
In the case of Carroll's work, the most internal words tended to be
more immediately related to the main thematic of the book, while the
most external nodes (borders) related to secondary concepts.

The effectiveness of the reported approach has paved the way to a
number of promising further investigations, including its
application for the identification of the borders in several
important real-world networks such as gene regulation, airports, and
anatomical networks~\cite{Viana2008}).  A particularly interesting
prospect would be to apply the methodology to detect the borders in
pictures where the objects do not have well-defined contours (e.g.
are composed by textures and clouds).

\section*{Acknowledgments}

Bruno A. N. Traven\c{c}olo is grateful to FAPESP for financial
support (07/02938-5), Matheus P. Viana thanks to FAPESP for
financial support (07/50882-9) and Luciano da Fontoura Costa thanks
to CNPq (301303/06-1) and FAPESP (05/00587-5) for financial support.
The authors also thank Lucas Antiqueira for building the network of
words from Carroll's book.

\section*{References}
\bibliographystyle{unsrt}
\bibliography{net_border}

\end{document}